\documentclass[prd,a4paper,twocolumn,superscriptaddress]{revtex4}
\usepackage{graphicx}
\usepackage{amsfonts}
 
\newcommand{\be}{\begin{equation}}
\newcommand{\ee}{\end{equation}}
\newcommand\lsim{\mathrel{\rlap{\lower4pt\hbox{\hskip1pt$\sim$}}
    \raise1pt\hbox{$<$}}}
\newcommand\gsim{\mathrel{\rlap{\lower4pt\hbox{\hskip1pt$\sim$}}
    \raise1pt\hbox{$>$}}}
\newcommand\esim{\mathrel{\rlap{\raise2pt\hbox{\hskip0pt$\sim$}}
    \lower1pt\hbox{$-$}}}

\begin{document}

\title{Evolution of the fine-structure constant in the non-linear regime}

\author{P.P. Avelino}
\email[Electronic address: ]{ppavelin@fc.up.pt}
\affiliation{Centro de F\'{\i}sica do Porto, Rua do Campo Alegre 687,
4169-007, Porto, Portugal}
\affiliation{Departamento de F\'{\i}sica da Faculdade de
Ci\^encias da Universidade do Porto, Rua do Campo Alegre 687,
4169-007, Porto, Portugal}
\author{C.J.A.P. Martins}
\email[Electronic address: ]{C.J.A.P.Martins@damtp.cam.ac.uk}
\affiliation{Centro de F\'{\i}sica do Porto, 
Rua do Campo Alegre 687, 4169-007, Porto, Portugal}
\affiliation{Department of Applied Mathematics and Theoretical
Physics, Centre for Mathematical Sciences,\\ University of
Cambridge, Wilberforce Road, Cambridge CB3 0WA, United Kingdom}
\author{J. Menezes}
\email[Electronic address: ]{jmenezes@fc.up.pt}
\affiliation{Centro de F\'{\i}sica do Porto, Rua do Campo Alegre 687,
4169-007, Porto, Portugal}
\affiliation{Departamento de F\'{\i}sica da Faculdade de
Ci\^encias da Universidade do Porto, Rua do Campo Alegre 687,
4169-007, Porto, Portugal}
\author{C. Santos}
\email[Electronic address: ]{cssilva@fc.up.pt}
\affiliation{Centro de F\'{\i}sica do Porto, Rua do Campo Alegre 687, 
4169-007, Porto, Portugal}
\affiliation{Departamento de F\'{\i}sica da Faculdade de
Ci\^encias da Universidade do Porto, Rua do Campo Alegre 687,
4169-007, Porto, Portugal}

\date{13 December 2005} 
\begin{abstract} 
We study the evolution of the fine-structure constant, $\alpha$, induced by non-linear density perturbations in the context of the simplest class of quintessence models with a non-minimal coupling to the electromagnetic field, in which the two available free functions (potential and gauge kinetic function) are Taylor-expanded up to linear order. We show that the results obtained using the spherical infall model for an infinite wavelength inhomogeneity are inconsistent with the results of a local linearized gravity study and we argue in favour of the second approach. We also discuss recent claims that the value of $\alpha$ inside virialised regions could be significantly different from the background one on the basis of these findings. 
\end{abstract} 
\maketitle 
 
\section{Introduction} 

One of the deepest question of modern physics is whether or not there are fundamental scalar fields in nature. For example, they are a key ingredient in the standard model of particle physics (cf. the Higgs particle, which is supposed to give mass to all other particles and make the theory gauge-invariant), but after several decades of accelerator experiments there is still no shred of experimental evidence for them.

The early universe is a much better (not to mention cheaper) laboratory for fundamental physics. Observations suggest that the recent universe is dominated by an energy component whose gravitational behaviour is quite similar to that of a cosmological constant (as first introduced by Einstein). This could of course be the right answer, but the observationally required value is so much smaller than what would be expected from particle physics that a dynamical scalar field is arguably a more likely explanation. Now, the slow-roll of this field (which is mandatory so as to yield negative pressure) and the fact that it is presently dominating the universe imply (if the minimum of the potential vanishes) that the field vacuum expectation value today must be of order $m_{Pl}$, and that its excitations are very light, with $m\sim H_0\sim 10^{-33}$ eV. But a further consequence of this is seldom emphasized \cite{CARROLL}: couplings of this field lead to observable long-range forces and time-dependence of the constant of nature (with corresponding violations of the Einstein Equivalence Principle).

Measurements of various dimensionless couplings, such as the fine-structure constant $\alpha$ (which will be the focus of this paper) or the electron to proton mass ratio \cite{THOMPSON} are therefore unique tests of fundamental physics. Note that since the scalar field is effectively massless on solar system scales, it should in principle be easier to find new physics on astrophysical and cosmological scales. Moreover, bounds on varying couplings restrict the evolution of the scalar field and provide constraints on dark energy \cite{PARKINSON,NUNES} that, with new datasets becoming available in the near future will be complementary to (and indeed more powerful and constraining than) those obtained by traditional means.

Let us now focus on the fine-structure constant, $\alpha$, which among other things measures the strength of the electromagnetic interaction. The good news is then that, since the standard physics is changed in a number of key ways if there is a spacetime variation of $\alpha$, there are many different ways in which measurements of $\alpha$ can be made. To name just a few, locally one can use atomic clocks \cite{Marion} or the Oklo natural nuclear reactor \cite{Fujii,Lamoreaux}. On the other hand, on astrophysical and cosmological scales a lot of work has been done on measurements using quasar absorption systems \cite{Webb1,Webb2,Murphy,Chand} and the cosmic microwave background \cite{Avelino1,Avelino2,Martins1,Martins2,Rocha}.

The bad news, however, is that these different measurements probe very different environments, and therefore it is not trivial to compare and relate them. Simply comparing at face value numbers obtained at different redshifts, for example, it is at the very least too naive, and in most cases manifestly incorrect. Indeed, detailed comparisons can often only be made in a model-dependent way, meaning that one has to specify a cosmological model (crucial to define a clock in the universe, that is, a timescale) and/or a specific model for the evolution of $\alpha$ as a function of redshift. Simply assuming, for example, that alpha grows linearly with time (so that its time derivative is constant) is not satisfactory, as one can easily see that no sensible particle physics model will ever yield such a dependence for any significant redshift range. Last but not least, new methods are being developed for measuring $\alpha$ using emission lines \cite{SDSS,JARLE,GERKE} as well as the electron to proton mass ratio (using absorption) \cite{Ivanchik1,Ivanchik2,Ubachs}, so the issue of detailed comparisons between datasets will be even more important for the next generation of datasets.

Here, we discuss one specific aspect of this issue. The scalar field responsible for the variation of $\alpha$ will (in any sensible particle physics model) 
couple to the matter sector. Among other things, this implies that when, in the course of the cosmological evolution, inhomogeneities grow, become non-linear and decouple from the background evolution, the same could happen to the local variations of $\alpha$. This has been previously studied in \cite{Barrow1,Mota1,Mota2} using a simple spherical infall model for the evolution of infinite wavelength density perturbations and a particular generalization of the Bekenstein model \cite{Sandvik} for the evolution of $\alpha$. (See also \cite{Barrow2} for a discussion of large scale variations of $\alpha$.) It was found that in the linear regime and in the matter era the variation of $\alpha$ would follow the density contrast. Moreover, it was also claimed that this approach was valid in the non-linear regime (meaning turnaround and collapse). Here we revisit these results, in particular questioning the applicability of the spherical infall model. Note that in the particular case of the models of \cite{Barrow1,Mota1,Mota2,Sandvik} it is enforced (purely by hand) that the evolution of $\alpha$ is driven by a coupling to charged non-relativistic matter alone and in that case the scalar field cannot also provide the dark energy. For that reason one should see these models as toy models that are useful for computational purposes. To some extent a similar comment applies to the models that we shall consider, though the reason here is purely their extreme simplicity. However, we do expect that at least at a qualitative level our results will be representative of more realistic models.

While this paper was being written up, ref. \cite{Shaw} appeared. This provides a more detailed and mathematically-inclined analysis of local variations in physical `constants', but does confirm our results. Our approach, while much simpler, has the advantage of making explicit the reasons why spatial variations have to be small, and why the use of the spherical collapse model of a infinite wavelength perturbation is inadequate.

We will start in Sect. \ref{lin} with a brief description of the models that we will be using. In Sect. \ref{nonlin} we discuss the non-linear evolution of the fine structure constant using two different approaches. Finally, we describe and discuss our results in Sect. \ref{res}, and present our conclusions in Sect. \ref{end}. Throughout this paper we shall use fundamental units with $\hbar=c=G=1$.
 
\section{\label{lin}The linearized models} 
 
We will consider the class of models where a neutral scalar field is non-minimally coupled (via a gauge kinetic function $B_F$) to electromagnetism, namely
\begin{equation} 
{\cal L} = {\cal L}_\phi + {\cal L}_{\phi F} + {\cal L}_{\rm other}\, , 
\end{equation} 
where
\begin{equation} 
{\cal L}_\phi= \frac{1}{2}\partial^\mu \phi \partial_\mu \phi - V(\phi)\, , 
\end{equation} 
\begin{equation} 
{\cal L}_{\phi F}= -\frac{1}{4}B_F(\phi) F_{\mu \nu} F^{\mu \nu}\,, 
\end{equation} 
and ${\cal L}_{\rm other}$ is the Lagrangian density of the other fields. 
We will make the simplifying assumption that both $V$ and $B_F$ are linear functions of $\phi$, namely
\begin{equation} 
V(\phi) = V(\phi_0) + \frac{dV}{d\phi} \left(\phi-\phi_0\right)\, , 
\end{equation} 
where $dV/d\phi$ is assumed to be a constant, the subscript `0' indicates 
that the variable local value is to be evaluated at the 
present time, and
\begin{equation} 
B_F (\phi) = 1-\zeta k (\phi-\phi_0)\, , 
\end{equation} 
where $k^2=8\pi$ and Equivalence principle tests require that 
\cite{Will,Damour}
\begin{equation} 
|\zeta| < 5\times10^{-4}\, . 
\end{equation} 
Since $B_F=\alpha_0/\alpha$, this implies that
\begin{equation} 
\frac{\alpha}{\alpha_0}= 1+\zeta k \left(\phi-\phi_0\right)\, , 
\end{equation} 
up to linear order.
The equation of motion for the field $\phi$ is given by 
\begin{equation} 
\label{phieq} 
\Box \phi=\frac{\zeta k}{4} F_{\mu \nu} F^{\mu \nu} - \frac{dV}{d\phi}\, . 
\end{equation}

Therefore the evolution of the average value of $\alpha$ with physical 
time is given approximately by
\begin{equation} 
\label{phieq3} 
{\ddot {\bar \alpha}}+3H{\dot {\bar \alpha}} \sim (- \xi_1 \rho_m + \xi_2){\bar \alpha}\,, 
\end{equation}
where we have defined $F_{\mu \nu} F^{\mu \nu}/4 = - \gamma_F \rho_m$, 
$\xi_1=\gamma_F (\zeta k)^2 < 6 \times 10^{-6} \gamma_F$ and $\xi_2=\zeta k (dV/d\phi)$.
On the other hand, linearizing eqn. (\ref{phieq}) to obtain a local static 
solution in a slightly perturbed Minkowski space we obtain 
\begin{equation} 
\label{phieq2} 
\frac{\nabla^2 \alpha}{\bar \alpha} \sim \xi_1 \delta \rho_m\, . 
\end{equation}
In the following, for simplicity we shall assume that $\xi_2=0$ and we will 
define $\xi=\xi_1$. Although a generalization to models with $dV/d\phi \neq 0$ 
is straightforward, our main results will not be dependent on this 
particular assumption.

\section{\label{nonlin}Non-linear evolution of $\alpha$} 

We will consider the evolution of the value of $\alpha$ due to the growth 
of a uniform spherical matter inhomogeneity with a final size $r_0$ using 
two different approaches. In what we shall refer to as the \textit{local approximation}, we solve the Poisson equation 
(\ref{phieq2}) outside the spherical distribution of mass, $M$, to 
determine the 
spatial variation of $\alpha$ as
\begin{equation} 
\label{phieq1} 
\frac{\delta \alpha}{\alpha}\equiv \frac{\alpha(r)-{\bar \alpha}}{\bar \alpha} 
\sim - \xi \frac{M}{4\pi r} \,,
\end{equation}
identifying ${\bar \alpha}$ with $\alpha(r=\infty)$. Given that  
$M=4\pi \delta \rho_m r^3/3$ we have
\begin{equation} 
\label{model1} 
\frac{\delta \alpha}{\alpha}=- \frac{1}{3} \xi \delta \rho_m r^2\,.
\end{equation}
Note that this calculation assumes that the spherical inhomogeneity has a 
radius significantly smaller than the horizon in order for eqn. (\ref{phieq2}) 
to be a valid approximation locally. 

A different calculation was introduced in \cite{Mota1,Mota2} 
where the evolution of a uniform (that is, \textit{infinite wavelength}) perturbation in  $\alpha$ was studied. The 
evolution of two homogeneous and isotropic universes was considered with 
the initial conditions set at some initial time $t_i$ deep in the matter 
era. Here, we shall consider a similar calculation where the 
background universe in a spatially flat Friedman-Robertson-Walker 
universe containing matter, radiation and a cosmological constant whose 
dynamics is described by
\begin{equation} 
\frac{\ddot a}{a}=-H_i^2\left[\frac{\Omega_{mi}}{2} \left(\frac{a}{a_i}\right)^{-3} 
+\Omega_{ri} \left(\frac{a}{a_i}\right)^{-4} -\Omega_{\Lambda i}\right]\, , 
\end{equation} 
with $\Omega_{ki}=1-\Omega_{mi}-\Omega_{ri}-\Omega_{\Lambda i}=0$ and 
taking the initial time $t_i$ to be deep into the radiation era. We also 
consider a perturbed closed universe with $H_i^P \sim H_i$, 
$\Omega^P_{ri}=\Omega_{ri}$, $\Omega^P_{\Lambda i}=\Omega_{\Lambda i}$, 
$\Omega^P_{mi}=\Omega_{mi}+\Delta \Omega_{mi}$, and $\Omega^P_{ki}=-
\Delta \Omega_{mi}$.
In this paper, for simplicity we include the contribution of the energy 
density of the field $\phi$ in the value of $\Omega_{\Lambda}$ thus 
neglecting the contribution of the kinetic contribution to 
energy density of the field $\phi$, which is nevertheless constrained to be 
small. Also, since any variation of the fine structure constant 
from the epoch of  nucleosynthesis onwards is expected to be very small 
\cite{Avelino2,Martins2} we neglect the minor contribution that such 
a variation has in the evolution of the baryon density (included in 
$\Omega_{m0}$).

In this approach the 
average values of $\alpha$ in the background and perturbed universes 
are computed using eqn. (\ref{phieq3}) so that 
\begin{equation} 
\label{phieq5} 
\left({\ddot \frac{\Delta \alpha}{\alpha}}\right)_{b,\infty}+3H\left({\dot \frac{\Delta \alpha}{\alpha}}\right)_{b,\infty} \sim -\xi 
\frac{3 \Omega_m H^2}{8\pi}\,,
\end{equation}
where
\begin{equation}
\frac{\Delta \alpha}{\alpha} = 
\frac{{\bar \alpha}-{\bar \alpha}_i}{{\bar \alpha}_i}
\end{equation}
and the subscripts $b$ and $\infty$ indicate that ${\bar \alpha}$ is 
calculated for the background and perturbed universes respectively.
By contrast, in the local approximation $\Delta \alpha/\alpha$ can be computed as  
\begin{equation}
\left(\frac{\Delta \alpha}{\alpha}\right)_\ell \sim 
\left(\frac{\Delta \alpha}{\alpha}\right)_b +\frac{\delta \alpha}{\alpha}
\end{equation}
where
\begin{equation}
\label{phieq6} 
\frac{\delta \alpha}{\alpha} \sim - \xi 
\frac{\left(H^2 \Omega_m\right)^P-\left(H^2 \Omega_m\right)}{8\pi H^2} 
\left(a^P r_0 H\right)^2\,.
\end{equation}
We can already anticipate that 
these two different approaches to estimate spatial variations of $\alpha$ 
will produce very different results since in the second approach there 
is no reference to the size of the fluctuation (an infinite wavelength 
approximation is considered). 

\section{\label{res}Results and discussion}

\begin{figure} 
\includegraphics[width=3.5in,keepaspectratio]{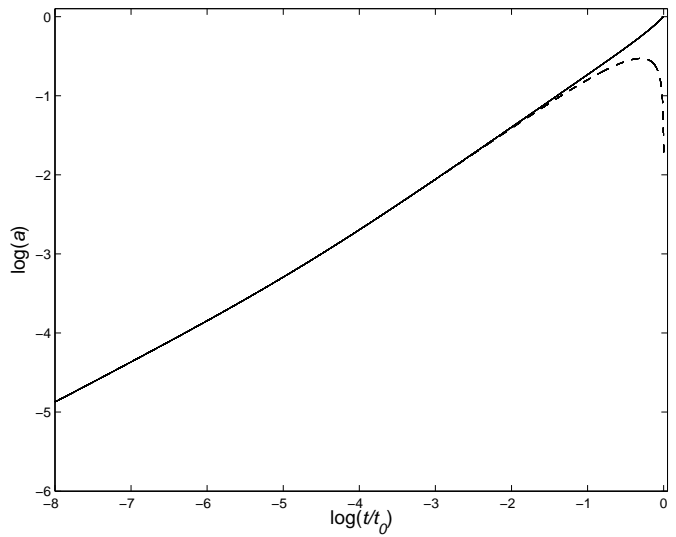} 
\caption{\label{fig1}Evolution of the scale factor, $a$, as a function of 
physical time, $t$, for the background and perturbed universes (solid and 
dashed lines respectively)} 
\end{figure} 

\begin{figure} 
\includegraphics[width=3.5in,keepaspectratio]{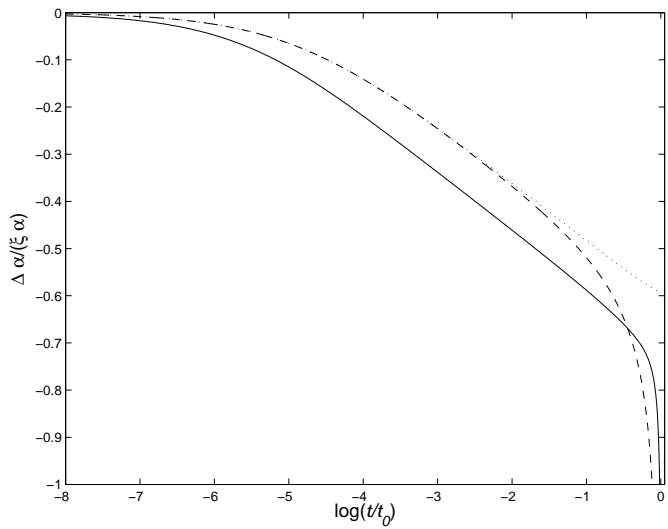} 
\caption{\label{fig2}Evolution of the fine structure constant, $\alpha$, 
as a function of physical time, $t$, in the vicinity of spherical 
distribution of mass including the spatial variations of $\alpha$ calculated 
using the local approximation with $r_0=2H_0^{-1}$ (solid line) and the infinite wavelength approximation (dashed line). The dotted line represents the background evolution of $\alpha$. Note that $\xi < 6 \times 10^{-6} \gamma_F$ is a 
very small number.}
\end{figure} 

\begin{figure} 
\includegraphics[width=3.5in,keepaspectratio]{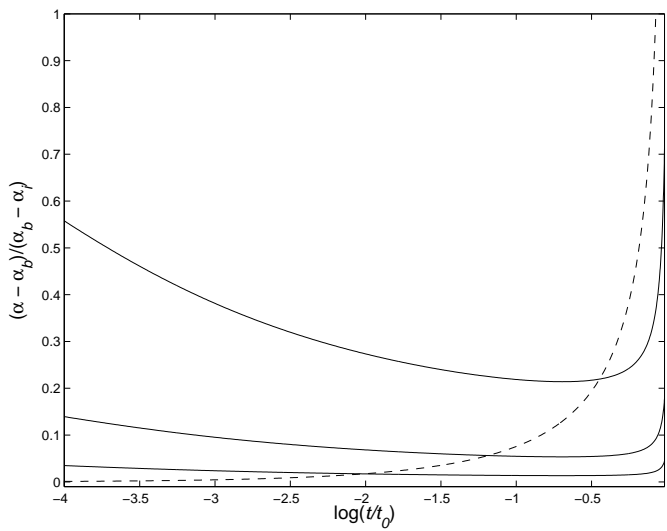} 
\caption{\label{fig3} Evolution of $(\alpha-\alpha_b)/(\alpha_b-\alpha_i)$, in various 
models as a function of physical time, $t$. The dashed line represents the 
infinite wavelength approximation while the solid lines use the local 
approximation with $r_0=2H_0^{-1}$, $r_0=H_0^{-1}$ and $r_0=H_0^{-1}/2$ 
(top to bottom respectivelly).}
\end{figure} 

The initial values of the various cosmological parameters were chosen in 
such a way that at the present time we have $\Omega_m^0=0.29$, 
$\Omega_{\Lambda_0}=0.71$ and $\Omega_{r0}=8.4 \times 10^{-5}$ 
\cite{Spergel} for the background universe. Also, the value of 
$\Delta \Omega_{mi}$ is such that the collapse of the perturbed 
universe occurs near the present epoch. This is clearly seen 
in Fig. 1 which plots the evolution of the scale factor, $a$ as a function 
of the physical time, $t$, for the background and perturbed universes 
(solid and dashed lines respectively). Note that, according to the spherical 
collapse model, a uniform density perturbation will become infinite in a 
finite time. Clearly this is not realistic (at least for local fluctuations) 
since in practice the collapse is stopped by virialisation.

In Fig. 2 we compare the background evolution of $\Delta \alpha/\alpha$ 
(dotted line) with that obtained using the two different models described 
in the previous section. While the evolution of $\Delta \alpha/\alpha$ 
obtained using the local approximation (dashed line) depends on the final size of the perturbation, $r_0$, no reference to the perturbation size is made 
in the second case (solid line) where an uniform, infinite wavelength was 
considered. Hence, the latter model will produce a single result 
while the first method will produce a result which depends explicitly 
on the final perturbation size, $r_0$. In Fig. 2 the results for 
the local approximation assume that $r_0=2H_0^{-1}$. This is clearly an unrealistic value 
of $r_0$ which we use in order to get a sizable depart from the background 
$\Delta \alpha/\alpha$ (note that in order to be self-consistent the model 
requires that $r \ll H^{-1}$). However, the result can easily be rescaled 
for realistic values of $r_0$ using (\ref{phieq6}) (for example for a 
typical size of a galaxy cluster).

In Fig. 3 we show the evolution of $(\alpha-\alpha_b)/(\alpha_b-\alpha_i)$, in various 
models as a function of physical time, $t$. The dashed line represents the 
infinite wavelength approximation while the solid lines use the local 
approximation with $r_0=2H_0^{-1}$, $r_0=H_0^{-1}$ and $r_0=H_0^{-1}/2$ 
(top to bottom respectivelly). We clearly see that as we move towards 
smaller (more realistic) values of $r_0$ the local approximation will 
give negligible variations of $\alpha$.

It is interesting to discuss the evolution of $\delta \alpha/\alpha$  
in the linear regime during the matter and radiation eras. Note that in 
the linear regime the physical radius of the spherical overdensity grows as
\begin{equation}
r \propto a\,,
\end{equation}
and therefore
\begin{equation}
M \propto \delta \rho_m a^3 = \delta_m {\bar \rho}_m a^3\,
\end{equation}
which is 
approximately constant during the radiation era and grows proportionally to $a$ 
during the matter era. Hence deep into the radiation era 
\begin{equation}
\frac{\delta \alpha}{\alpha} \propto a^{-1}\,,
\end{equation}
while deep into the matter era
\begin{equation}
\frac{\delta \alpha}{\alpha} \sim  {\rm constant}\,.
\end{equation}
It is also important to point out 
that the amplitude of the $\alpha$ variation (for a given $\delta \rho_m$) 
depends quadratically on the radius of the spherical overdensity 
($\Delta \alpha/\alpha \propto r^2$). Consequently, for $r_0 \ll H_0^{-1}$ 
the spatial variations predicted by the local approximation will be very small. The background evolution of $\Delta \alpha/\alpha$ in linearized Bekenstein 
models has been previously studied in \cite{Linear} and it is well known that 
$\Delta \alpha/\alpha \to 0$ deep into the 
radiation era.

This is confirmed by the results plotted in Fig. 2 which clearly 
confirm the above discussion. In both approximations
\begin{equation}
\frac{\Delta \alpha}{\alpha} \to \left(\frac{\Delta \alpha}{\alpha}\right)_b 
\to 0\,,
\end{equation}
deep into the radiation era. On the other hand  
\begin{equation}
\left(\frac{\Delta \alpha}{\alpha}\right)_\ell- 
\left(\frac{\Delta \alpha}{\alpha}\right)_b \sim {\rm constant}\,,
\end{equation}
deep into the matter era. Also, as expected the evolution of 
$\Delta \alpha/\alpha$ in the infinite wavelength approximation only departs significantly from the background one near the present time (when the collapse of the perturbed universe occurs).

\section{\label{end}Conclusions}

We have presented and contrasted two different approaches for the 
determination of the non-linear evolution of $\alpha$. While in the local approximation 
we assume that the size of the perturbation which gives rise to spatial 
variation of $\alpha$ is much smaller than the Hubble radius, $H^{-1}$, 
in the infinite wavelength approximation the opposite is required for self-consistency. Hence, it is not surprising 
that the results obtained using both models are so different. The 
crucial question is which one provides the right answer when 
applied to cosmology. Here, we argue that local approximation provides the right answer since the non-linear effects are only expected to be important 
in this context on scales much smaller than $H_0^{-1}$ (scales smaller 
than the a typical galaxy cluster size). On such small scales, we therefore predict that the spatial variations of $\alpha$ generated in 
the simplest models should be too small to be of any cosmological interest.
These results confirm that it is difficult to generate significant 
large-scale spatial variations of $\alpha$ \cite{Menezes1, Menezes2, Barrow2} 
even when we account for the evolution of the fine structure constant in the 
non-linear regime.

\begin{acknowledgments}
This work was funded by FCT (Portugal), through grant POCTI/CTE-AST/60808/2004, in the framework of the POCI2010 program, supported by FEDER.
J. Menezes was supported by a Brazilian Government grant - CAPES-BRAZIL 
(BEX-1970/02-0). 
\end{acknowledgments} 

\bibliography{alphanl}
\end{document}